\documentclass[twocolumn]{aastex63}
\usepackage{graphics,epsf}
\usepackage{amsmath}                
\usepackage{amsfonts}               
\usepackage{amssymb}                
\usepackage{epsfig}                 
\usepackage{appendix}
\usepackage{graphicx}
\usepackage{float}
\usepackage{color}
\usepackage{multirow}
\usepackage{colortbl}
\usepackage[para,online,flushleft]{threeparttable}

\hypersetup{citecolor=blue, 
            linkcolor=red, 
            menucolor=blue, 
            urlcolor=blue}  

\newcommand{\cm}{{~\rm cm}}
\newcommand{\m}{{~\rm m}}
\newcommand{\km}{{~\rm km}}
\newcommand{\s}{{~\rm s}}
\newcommand{\h}{{~\rm h}}

\newcommand{\g}{{~\rm g}}
\newcommand{\G}{{~\rm G}}
\newcommand{\K}{{~\rm K}}

\newcommand{\yr}{{~\rm yr}}

\begin{document}

\title{Pre-supernova outbursts by core magnetic activity}


\author[0009-0001-0720-6816]{Tamar Cohen}
\affiliation{Department of Physics, Technion, Haifa, 3200003, Israel; tamarco@campus.technion.ac.il; soker@physics.technion.ac.il}

\author[0000-0003-0375-8987]{Noam Soker}
\affiliation{Department of Physics, Technion, Haifa, 3200003, Israel; tamarco@campus.technion.ac.il; soker@physics.technion.ac.il}

\begin{abstract}
We conduct one-dimensional stellar evolutionary numerical simulations under the assumption that an efficient dynamo operates in the core of massive stars years to months before core collapse and find that the magnetic activity enhances mass loss rate and might trigger binary interaction that leads to outbursts. We assume that the magnetic flux tubes that the dynamo forms in the inner core buoy out to the outer core where there is a steep entropy rise and a molecular weight drop. There the magnetic fields turn to thermal energy, i.e., by reconnection. We simulate this energy deposition where the entropy steeply rises and find that for our simulated cases the envelope radius increases by a factor of $\simeq 1.2-2$ and luminosity by about an order of magnitude. These changes enhance the mass loss rate. The envelope expansion can trigger a binary interaction that powers an outburst. Because magnetic field amplification depends positively on the core rotation rate and operates in cycles, not in all cases the magnetic activity will be powerful enough to change envelope properties. Namely, only a fraction of core-collapse supernovae experience pre-explosion outbursts.  
\end{abstract}

\section{Introduction} 
\label{sec:intro}

The progenitors of some core-collapse supernovae (CCSNe) experience one or more pre-explosion outbursts (PEOs) before they explode. The PEOs lead to brightening and/or enhanced mass loss rate and take place mainly somewhere between tens of years to a few days before the explosion (e.g., \citealt{Foleyetal2007, Pastorelloetal2007, Smithetal2010, Mauerhanetal2013, Ofeketal2013, Pastorelloetal2013, Marguttietal2014, Ofeketal2014, SvirskiNakar2014, Tartagliaetal2016, Yaronetal2017, Wangetal2019, Bruchetal2020, Prenticeetal2020, Strotjohannetal2021}). The enhanced mass loss rate episode might even take place much earlier, i.e., during the core carbon-burning phase (e.g., \citealt{Moriyaetal2014, Marguttietal2016}).

Extra (to the regular nuclear burning) energy deposition to the envelope can lead to its expansion without a substantial mass loss (e.g., \citealt{McleySoker2014, LeungFuller2020}), or possibly with moderate to substantial mass ejection (e.g., \citealt{Quataertetal2016, Fuller2017, FullerRo2018, Tsangetal2022}). Several extra energy sources might lead to pre-explosion outbursts in CCSN progenitors. We list some here.   

(1) \textit{Pair instability.} There is the six-decades-old pair-instability mechanism \citep{RakavyShaviv1967, Barkatetal1967} that can lead to a huge mass loss before the explosion (e.g., \citealt{Chenetal2014}). This mechanism is limited to very massive stars that are very rare. 

(2) \textit{Core-excited waves.} In this process that was proposed and developed by \cite{QuataertShiode2012} and \cite{ShiodeQuataert2014} the vigorous convection in the core during the last stages of nuclear burning excites g-waves that propagate outward, turn to p-waves in the envelope, and deposit most of their energy in the outer layers of the envelope. The energy deposition leads to the PEO (for further studies see, e.g., \citealt{RoMatzner2017, WuFuller2021, WuFuller2022a}). 

(3) \textit{Magnetic activity in the core.} A dynamo might operate inside the core of pre-collapse stars (e.g., \citealt{Hegeretal2005, Wheeleretal2015, Zilbermanetal2018}). 
\cite{Peresetal2019} argue that the radiative zone above the iron core in pre-collapse cores of massive stars can store strong magnetic fields, up to $B\simeq 10^{13} \G$. Magnetic flux tubes with stronger magnetic fields buoy outward until they reach a steep entropy and/or molecular weight ($\mu$) gradient. This is the subject of this study. In that respect, we differ from \cite{SokerGilkis2017} who assumed that the magnetic flux tubes buoy all the way to the envelope. They further speculated that the
core convective shells power magnetic activity at $\approx 0.001$ times the convective luminosity. The magnetic energy injection to the envelope causes envelope expansion. A binary companion is required to power an energetic outburst (e.g., \citealt{McleySoker2014}). \cite{SokerGilkis2017} noted that magnetic activity requires minimum core rotation and that the
stochastic magnetic activity is on its high phase, conditions that are obeyed only in some CCSN progenitors, as observations require because not all CCSN progenitors experience PEOs. For example, SN 1987A did not experience a PEO. In some other cases, the presence of a compact circumstellar material (CSM) might be due to an extended envelope (e.g., \citealt{Dessartetal2017, Moriyaetal2017, Moriyaetal2018, Soker2021effer}) rather than a result of a PEO. 
An example is SN 2023ixf which had a compact CSM at explosion (e.g., \citealt{Bergeretal2023,  Bostroemetal2023, Grefenstetteetal2023, JacobsonGalanetal2023, Kilpatricketal2023, Koenig2023, Niuetal2023, Sarmah2023, SinghTejaetal2023, SmithNetal2023}) but no observed PEO (e.g., \citealt{DongYetal2023, Flinneretal2023, Jencsonetal2023, Neustadtetal2023, Panjkovetal2023, Soraisametal2023}). A possible explanation is that the CSM of SN 2023ixf is an effervescent zone \citep{Soker2023ixf}

The magnetic activity in the core with energy deposition inside the core rather than in the envelope is the subject of our study. We consider binary interaction as a substantial second extra energy source that removes most of the mass.   

(4) \textit{Binary interaction.} 
\cite{Chevalier2012} suggests a common envelope evolution with a neutron star or a black hole that leads to enhanced mass loss and to the explosion. In that scenario, the common envelope triggers the mass loss, eventually leading to the explosion. In a different setting of binary interaction, the activity starts with envelope expansion, either due to core activity (\citealt{Soker2013PEO, WuFuller2022Let}) or envelope hydrodynamical instability (e.g., \citealt{SmithArnett2014}). The envelope expansion triggers the binary interaction. No common envelope evolution is required in this case, only a strong binary interaction, possibly with jets that the companion launches (e.g., \citealt{DanieliSoker2019}). 
As well, the companion can be any type of star, compact or not. 

We describe our numerical scheme of energy injection in the outer core in section \ref{sec:Numerical}. We describe the influence of this energy injection on the star and the possible triggering of binary interaction in section \ref{sec:Results}.   
We summarize in section \ref{sec:Summary}. 

\section{Numerical procedure} 
\label{sec:Numerical}

We use the stellar evolution code \textsc{mesa} (Modules for Experiments in Stellar Astrophysics; \citealt{Paxtonetal2011, Paxtonetal2013, Paxtonetal2015, Paxtonetal2018, Paxtonetal2019, 2023ApJS..265...15J} )  from star formation to core-collapse.  We evolve a star with a zero-age main sequence (ZAMS) mass of $M_{\rm ZAMS}=12M_{\odot}$ and metallicity of $Z=0.02$.
 This mass is in the lower range of iron CCSNe as we are not aiming at electron-capture CCSNe. 
At the age of $t = 1.85\times 10^7 \yr$ the model reaches a radius of $R_{\ast} \simeq 830 R_{\odot}$, a luminosity of $L_{\ast} \simeq 5.8\times 10^4 L_{\odot}$ and has $\simeq 40 \%$ oxygen by mass in its core. For our goals, the important core property is that the oxygen burning excites strong convection in the center of the core and that this occurs about two years before the explosion. We turn off the mass loss during the evolution to obtain a star of $12 M_\odot$ at the relevant phase to us.
 We include 22 isotopes as the default of \textsc{mesa}. We note that \cite{Farmeretal2016} conclude that there is a need for $\simeq 130$ isotopes to reach convergence to $\simeq 10\%$ accuracy before core-collapse, e.g., in the location of the burning shells. We estimate that our qualitative results and our claims hold even for simulations with 22 isotopes.

\begin{figure}[]
	\centering
\includegraphics[trim=0cm 2.4cm 0.0cm 4.0cm ,clip, scale=0.43]{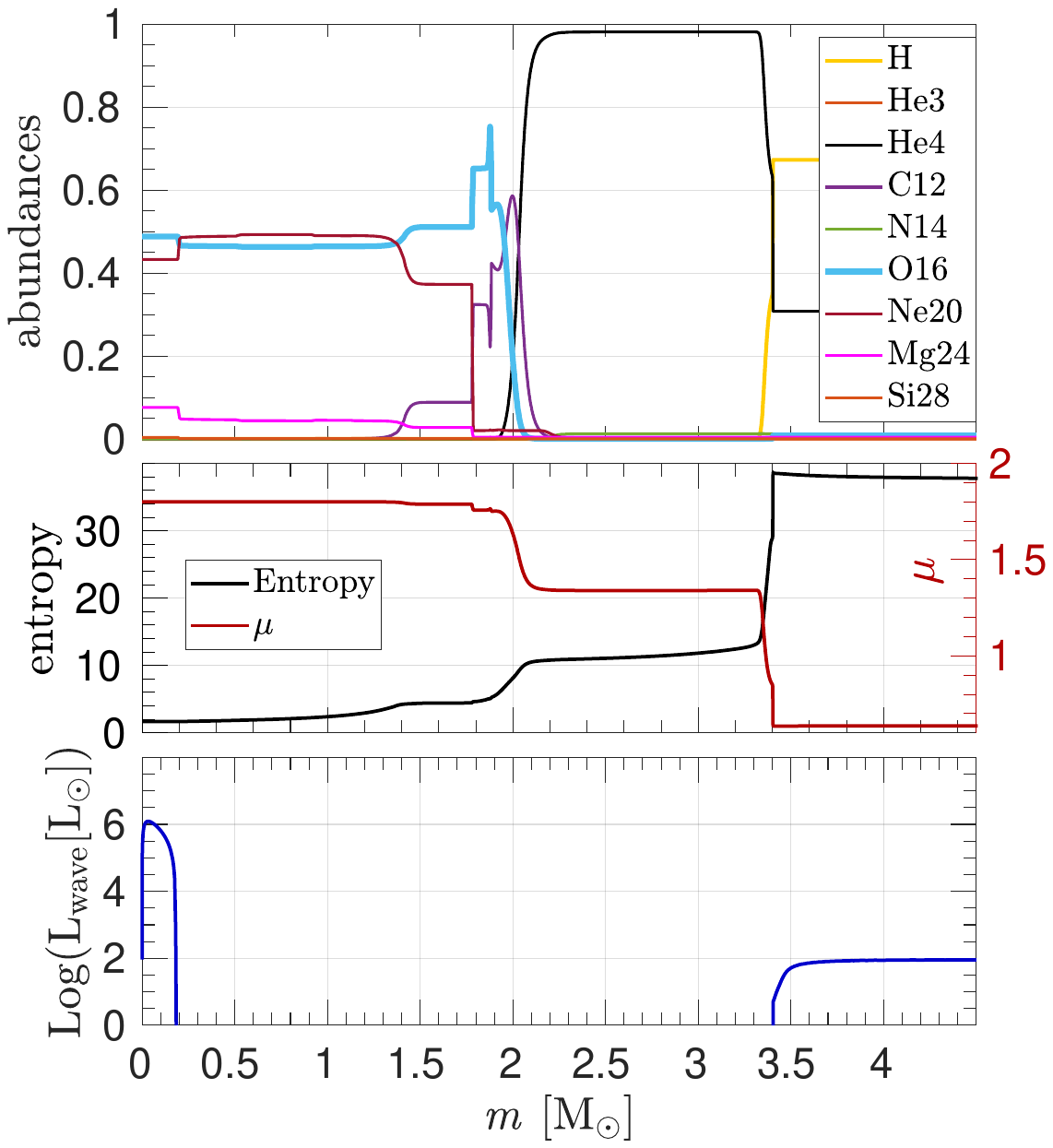}
 \\ 
\caption{The structure of the core and inner envelope of the stellar model we use here with  $M_{\rm ZAMS}=12M_{\odot}$ at $t = 1.85\times 10^7 \yr$ when $R_{\ast} \simeq 830 R_{\odot}$ and $L_{\ast} \simeq 5.8\times 10^4 L_{\odot}$. We turn off mass loss during the evolution. The upper panel presents the composition (fraction by mass), the middle panel presents the entropy (black line) and molecular weight (red line) while the lower panel presents the luminosity of the wave as calculated by equation (\ref{eq:LwaveFrac}).    
}
\label{fig:Profiles}
\end{figure}

We assume that the power of the magnetic activity in the core is like the wave luminosity as \cite{ShiodeQuataert2014} calculate (for details see \citealt{SokerGilkis2017}). This wave energy is \citep{LecoanetQuataert2013}, 
\begin{eqnarray}
\begin{aligned}
L_{\rm wave,0} & \approx \mathcal{M}^{5/8} L_{\rm conv} 
 = 
6 \times 10^5 
\\ & \times \left( \frac{\mathcal{M}}{0.001} \right) ^{5/8}
\left( \frac{L_{\rm conv}}{4.5 \times 10^7 L_\odot} \right)  L_\odot ,
\label{eq:LwaveFrac}
\end{aligned}
\end{eqnarray}
where 
\begin{equation}
L_{\rm conv} (r) = 4 \pi r^2 v^3_{\rm conv}(r) \rho(r)
\label{eq:Lconv} 
\end{equation}
is the luminosity that the convection carry at radius $r$, $\mathcal{M} = v_{\rm conv}/c_{\rm s}$ is the Mach number of the convective motion of velocity $v_{\rm conv}$, and $c_{\rm s}$ is the sound speed. The maximum value of the wave power in the core of our model is  $L_{\rm wave} =  6.05 \times 10^5 L_\odot$ at $m=0.07M_{\odot}$ where $r=2.1\times 10^{-3} R_{\odot}$, $v_{\rm conv}=4.09 \km \s^{-1}$, $\mathcal{M} =0.001$, and $L_{\rm conv} =4.53 \times 10^7 L_\odot $. 
 In Fig. \ref{fig:Profiles} we present some quantities as a function of the mass coordinate in the core of the evolved stellar model that we described above.  

We deposit energy into a shell with inner and outer boundaries that vary between the different cases that we simulate. The energy deposition power per unit mass in this shell is uniform. 

The magnetic activity scenario that \cite{SokerGilkis2017} proposed assumes that the core convection results in a dynamo that amplifies magnetic fields and forms magnetic flux tubes that buoy out. While \cite{SokerGilkis2017} assume that the magnetic flux tubes buoy all the way into the envelope, we here assume that the magnetic flux tubes stop rising at the steep entropy jump and the steep decrease in the molecular weight (e.g., \citealt{Peresetal2019}). 
 The steep entropy jump at $m \simeq 3.38 M_\odot$ corresponds to the interface between helium and hydrogen, namely, where hydrogen burns. In general, each steep entropy rise corresponds with a burning shell, i.e., an interface between elements (Fig. \ref{fig:Profiles}). This behavior is found in other stellar models at late evolutionary phases, e.g., the $M_{\rm ZAMS}=15 M_\odot$ and $M_{\rm ZAMS}=25 M_\odot$ models simulated by \cite{Peresetal2019}.

In the model, we simulate here the steep entropy jump and the large decrease in the molecular weight occurring at $m \simeq 3.38 M_\odot$ where the radius is $r(3.38) \simeq 2.5R_{\odot}$ and the density is $\rho (3.38) \simeq 7\times 10^{-4} \g \cm^{31}$.

This is the region where we deposit the energy that we assume to be magnetic energy. We also examine cases where the magnetic flux tubes from the core stop buoy at the inner and shallower entropy and molecular weight changes, i.e., at about $m=2M_\odot$, where $r(2) \simeq 4.6\times 10^{-2}R_{\odot}$ and $\rho (2) \simeq 9.5\times 10^{2} \g \cm^{31}$ (see Fig. \ref{fig:Profiles}).

There are studies of pre-explosion hydrodynamical mass ejection that are driven by energy deposition at the base of the envelope. \cite{KuriyamaShigeyama2020} conduct a thorough study of energy deposition at the base of the envelope using hydrodynamical code. \cite{Morozovaetal2020} inject the energy at the base of the hydrogen-rich envelope where the density is $\approx 10^{-5} \g \cm^{-3}$. \cite{Koetal2022} inject energy at a radius of $r \approx 1.7 \times 10^{12} \cm$. 
We differ from these studies in using a hydrostatic code and in injecting the energy at deeper zones, i.e., at the large entropy jump that is in the outer zone of the core and where we expect the magnetic flux tube to stop their buoyancy. 
\cite{Linialetal2021} use polytropic stellar models to study explosive mass ejection from the envelope. We differ in studying non-explosive mass injection. 
We differ from these four studies in attributing a major role to a binary companion in the mass ejection process (see discussion in section \ref{sec:intro}). The energy we deposit only causes stellar expansion, as we show in section \ref{sec:Results}. 

\section{Results} 
\label{sec:Results}

We present the results of energy deposition to the core of the $M_{\rm ZAMS}=12 M_{\odot}$ stellar model a few years before collapse as we described in section \ref{sec:Numerical}. 
In one type of simulations, we keep the energy-deposition power at $L_B=10^6 L_{\odot}$ and vary the shell into which we deposit the energy that mimics magnetic activity. The energy-deposition shell is around the jump in entropy at the moment when energy deposition starts $m=3.3-3.48 M_\odot$. 
There are four cases with energy deposition shell of $[3.3-3.48]M_\odot$ termed `Wide',  $[3.4-3.404]M_\odot$ `Narrow', $[3.35-3.4]M_\odot$ `Inner', and $[3.4-3.43]M_\odot$ `Outer'. In the following figures, we will mark these shells with a grey strip.  
In another set of simulations, we deposit energy at different powers in the range of $L_B=(0.333-1.2) \times 10^6 L_{\odot}$, all into a shell which is the Inner + Narrow shells defined above, i.e., $[3.35-3.404]M_{\odot}$).

\begin{figure}[]
	\centering
\includegraphics[trim=1.0cm 4.4cm 0.0cm 5.4cm ,clip, scale=0.42]{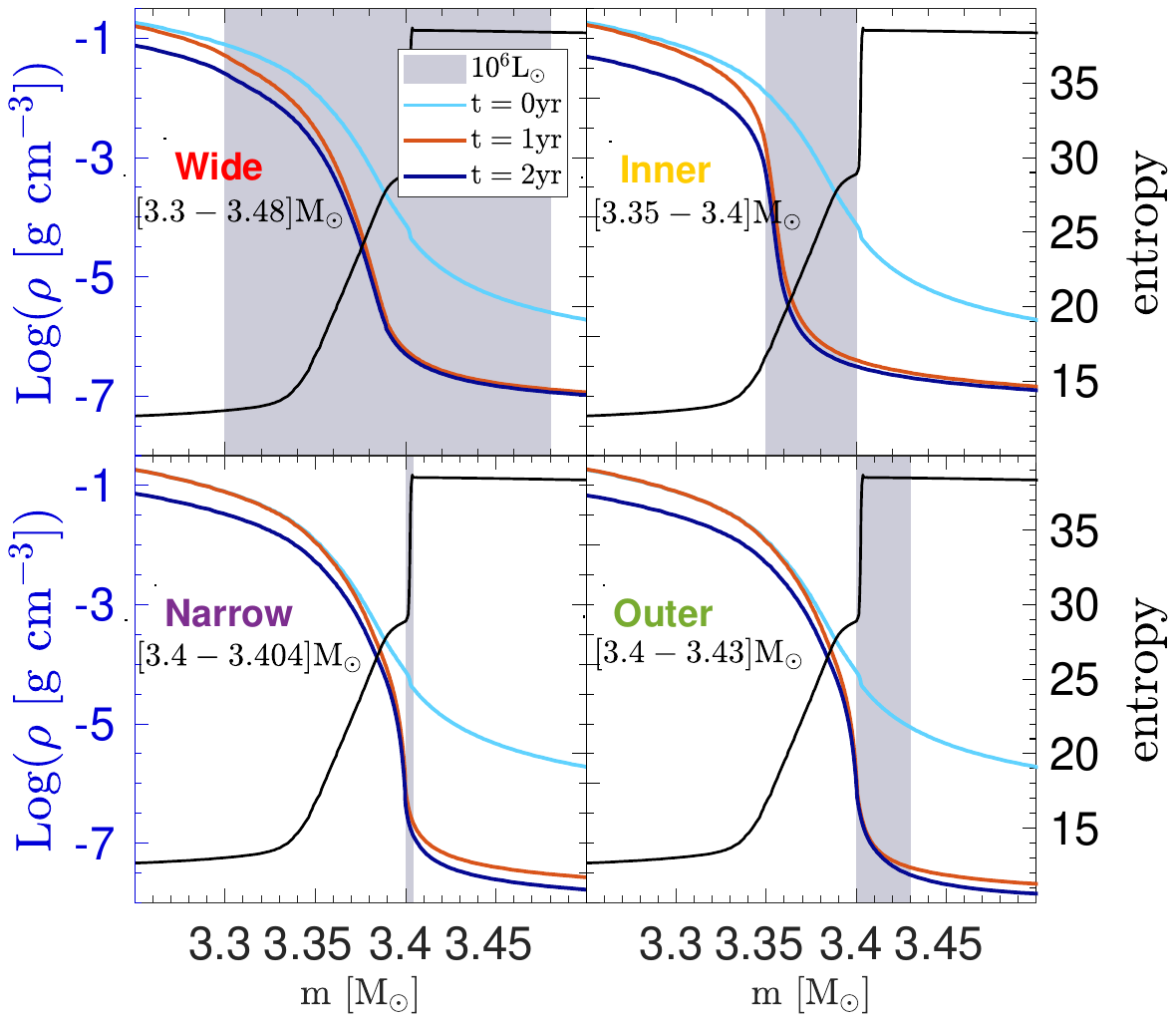}
 \\ 
\caption{ The stellar density profiles just before the energy deposition (cyan), one year into energy deposition (orange), and two years into the energy deposition which is a few days before core collapse (blue). The grey strips present the mass shell into which we deposit the energy that mimics magnetic activity. These are the cases with wide shell ($[3.35-3.404M_{\odot}]$), narrow shell ($[3.35-3.404M_{\odot}]$), inner shell ($[3.35-3.404M_{\odot}]$), and outer shell ($[3.35-3.404M_{\odot}]$). In all cases, the total power of energy deposition is $10^6 L_{\odot}$, and the deposition rate per unit mass is uniform. The black line is the initial entropy profile with the scale on the right. 
}
\label{fig:LogrhoEntropyVsMass4casesDiffmass}
\end{figure}

\begin{figure}[]
	\centering
\includegraphics[trim=0.0cm 2.70cm 0.0cm 3.8cm ,clip, scale=0.42]{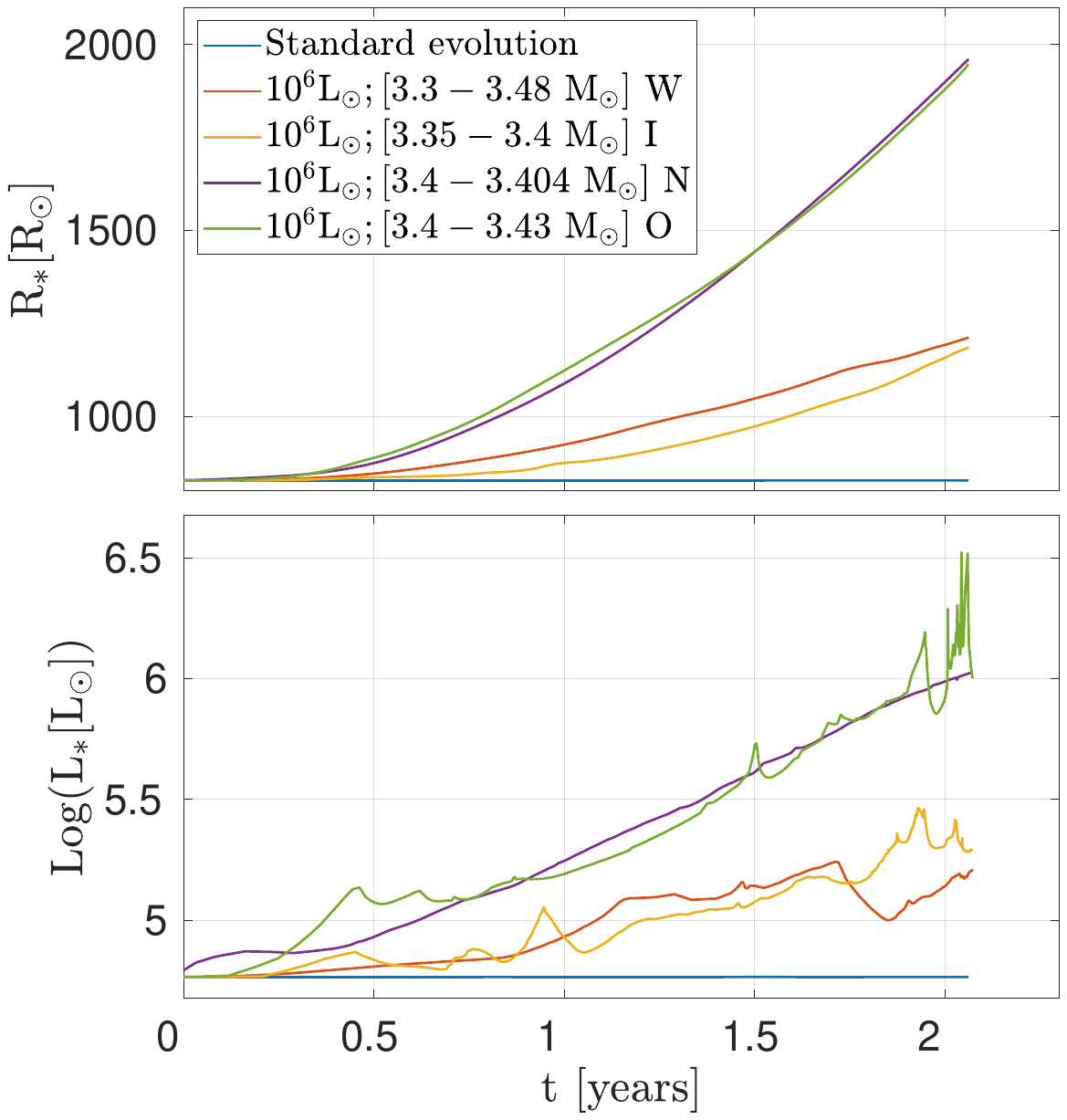}
 \\ 
\caption{ The stellar radius (upper panel) and stellar luminosity (lower panel) as a function of time for simulated cases that we present in Fig. \ref{fig:LogrhoEntropyVsMass4casesDiffmass}.  The time $t=0$ corresponds to the beginning of energy deposition that mimics magnetic activity in the core. 
}
\label{fig:RadiusLum4cases}
\end{figure}

\begin{figure}[]
	\centering
\includegraphics[trim=0.0cm 1.9cm 0.0cm 1.5cm ,clip, scale=0.42]{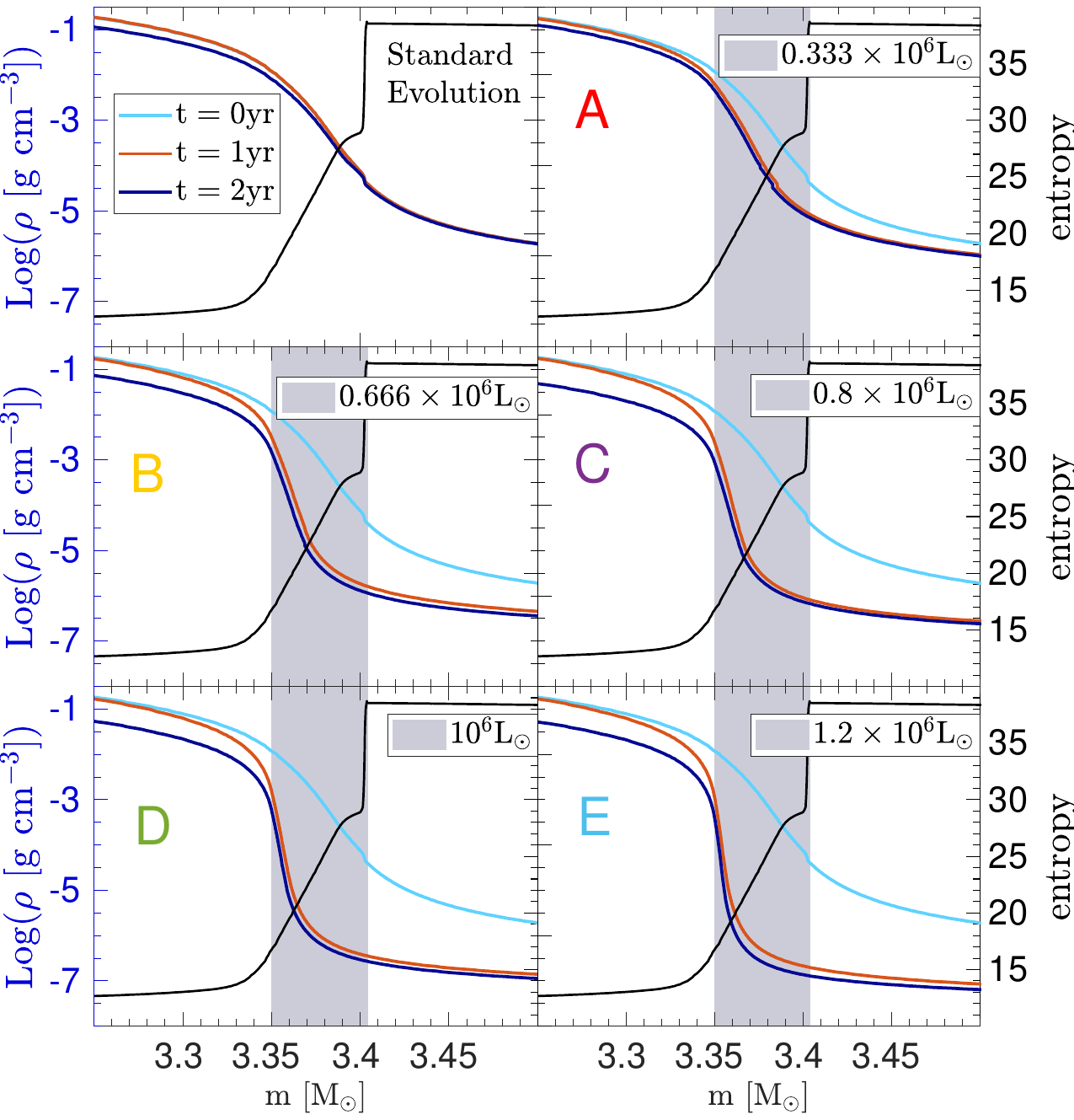}
 \\ 
\caption{ The stellar density profiles just before the energy deposition (cyan), one year into energy deposition (orange), and two years into the energy deposition, which is a few days before the core collapse (blue). 
The upper left panel presents the case without energy deposition. The other panels present cases with deposition energy in a shell of boundaries ($[3.35-3.404M_{\odot}]$), namely, the `narrow+inner' shells (grey; see Fig. \ref{fig:LogrhoEntropyVsMass4casesDiffmass}) and different powers as indicated in the insets. The black line is the initial entropy profile with the scale on the right. 
}
\label{fig:DepositionShells6case}
\end{figure}

\begin{figure}[]
	\centering
\includegraphics[trim=0.0cm 2.5cm 0.0cm 4.0cm ,clip, scale=0.42]{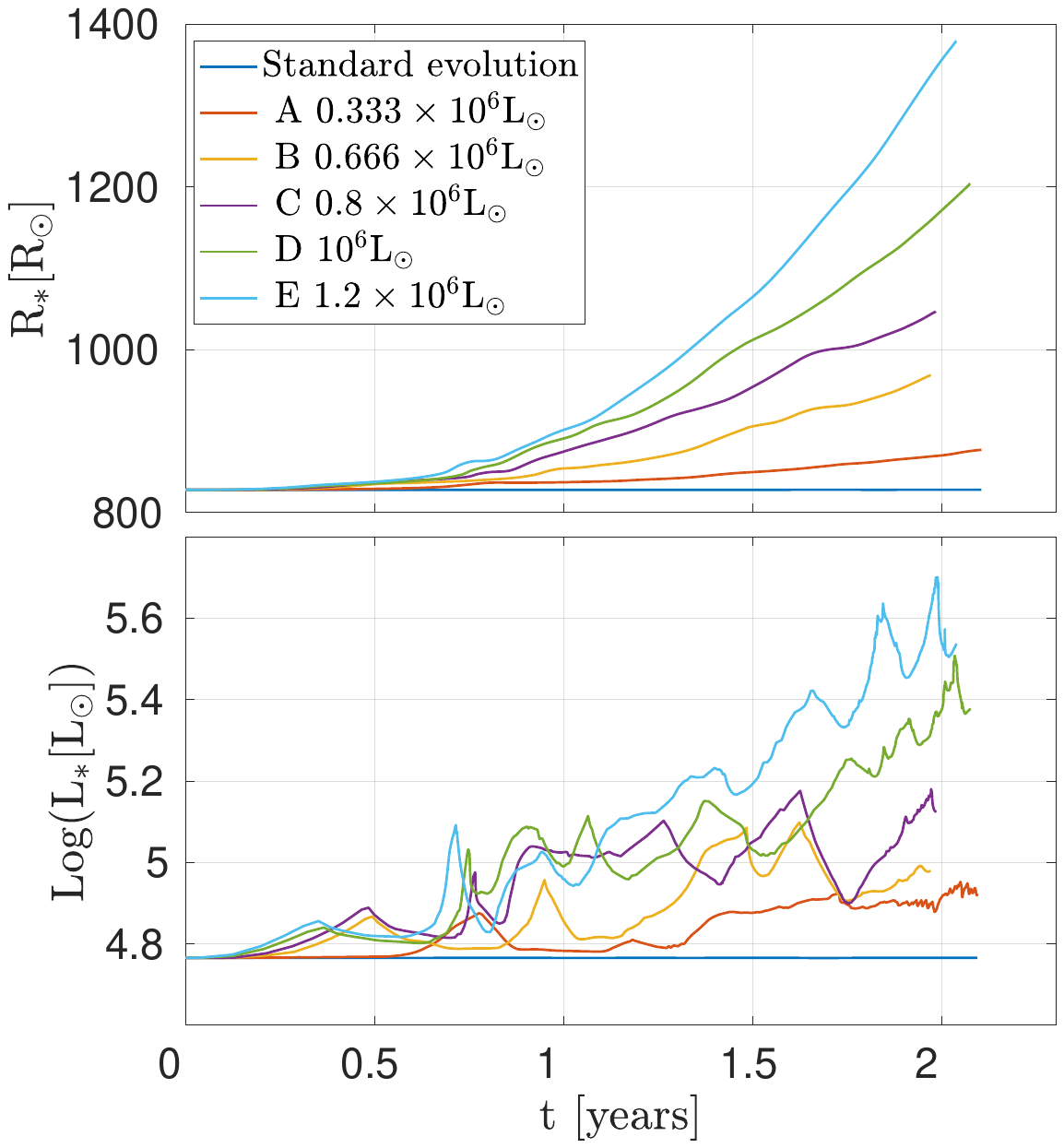}
 \caption{ Similar to Fig. \ref{fig:RadiusLum4cases} but for the six cases that we present in Fig. \ref{fig:DepositionShells6case}. 
}
\label{fig:RadiusLum6cases}
\end{figure}

In Figure \ref{fig:LogrhoEntropyVsMass4casesDiffmass} we present the density profiles in the core at three times for the first set of simulations. The times are when we start energy deposition, a year later, and two years later (in cases where the core collapses earlier, we present the density profile at the core-collapse time).  We also present the entropy profile at $t=0$ as we found that it does not change much along the evolution. 
In Figure \ref{fig:RadiusLum4cases} we present the evolution with time of the stellar radius and luminosity for the same cases. The graphs end when the core collapse starts. Similarly, we present the results for the set of simulations with varying energy deposition power in  Figures \ref{fig:DepositionShells6case}
and \ref{fig:RadiusLum6cases}. 

\section{Discussion and Summary} 
\label{sec:Summary}
We conducted one-dimensional stellar evolutionary simulations of a stellar model with $M_{\rm ZAMS}=12M_{\odot}$ to oxygen burning about 2-3 years before core collapse (section \ref{sec:Numerical}). 
We assume that magnetic activity in the core amplifies magnetic fields that buoy outward to the radius where entropy sharply increases and molecular weight rapidly decreases, around $m=3.38M_\odot$ (Figure \ref{fig:Profiles}). There, we assume, the magnetic energy is channeled to thermal energy. We mimicked the effect of magnetic energy by injecting energy to a shell around the entropy jump until the core collapses. We varied the power of the magnetic activity and the inner and outer boundaries of the shell (values in the insets of Figures \ref{fig:LogrhoEntropyVsMass4casesDiffmass} - \ref{fig:RadiusLum6cases}).  

We simulated 1D stellar models. Future studies should include 3D simulations to explore the full convection properties. We estimate the qualitative results would stay the same.   

Our main findings are as follows. (1) As we deposit energy to the relevant mass shell, the density in that shell decreases (Figures \ref{fig:LogrhoEntropyVsMass4casesDiffmass}  and \ref{fig:DepositionShells6case}) and the star expands (Figures \ref{fig:RadiusLum4cases} and \ref{fig:RadiusLum6cases}). (2) The factor by which the star expands is sensitive to the mass and exact location of the shell into which we deposit the energy; the inner boundary of the shell must be below the sharp entropy jump and the outer boundary must be above it. The envelope expansion increases as we decrease the mass of the shell and the envelope moves outwards. (3) The star might expand by tens of percent and up to by a factor of $\simeq 2$ for our simulated values. Namely, from a radius of $830 R_{\odot}$ when we start energy deposition to $\simeq (1200-2000)R_{\odot}$ just before collapse. (4) The stellar luminosity substantially increases from $6 \times 10^4 L_{\odot}$ to $\simeq(10^5-10^6)L_{\odot}$, with a larger increase for narrow and outer shells. (5) As we increase energy deposition power so do the stellar expansion (Figure \ref{fig:DepositionShells6case}) and its luminosity (Figure \ref{fig:RadiusLum6cases}). 

\cite{QuataertShiode2012} and \cite{ShiodeQuataert2014} suggested that the vigorous convection in the core excites waves that deposit their energy in the outer envelope. 
The magnetic activity can take place together with wave activity or be an alternative to wave activity and cause envelope expansion by itself. 

\begin{figure}[]
	\centering
\includegraphics[trim=0.0cm 2.9cm 0.0cm 4.0cm ,clip, scale=0.40]{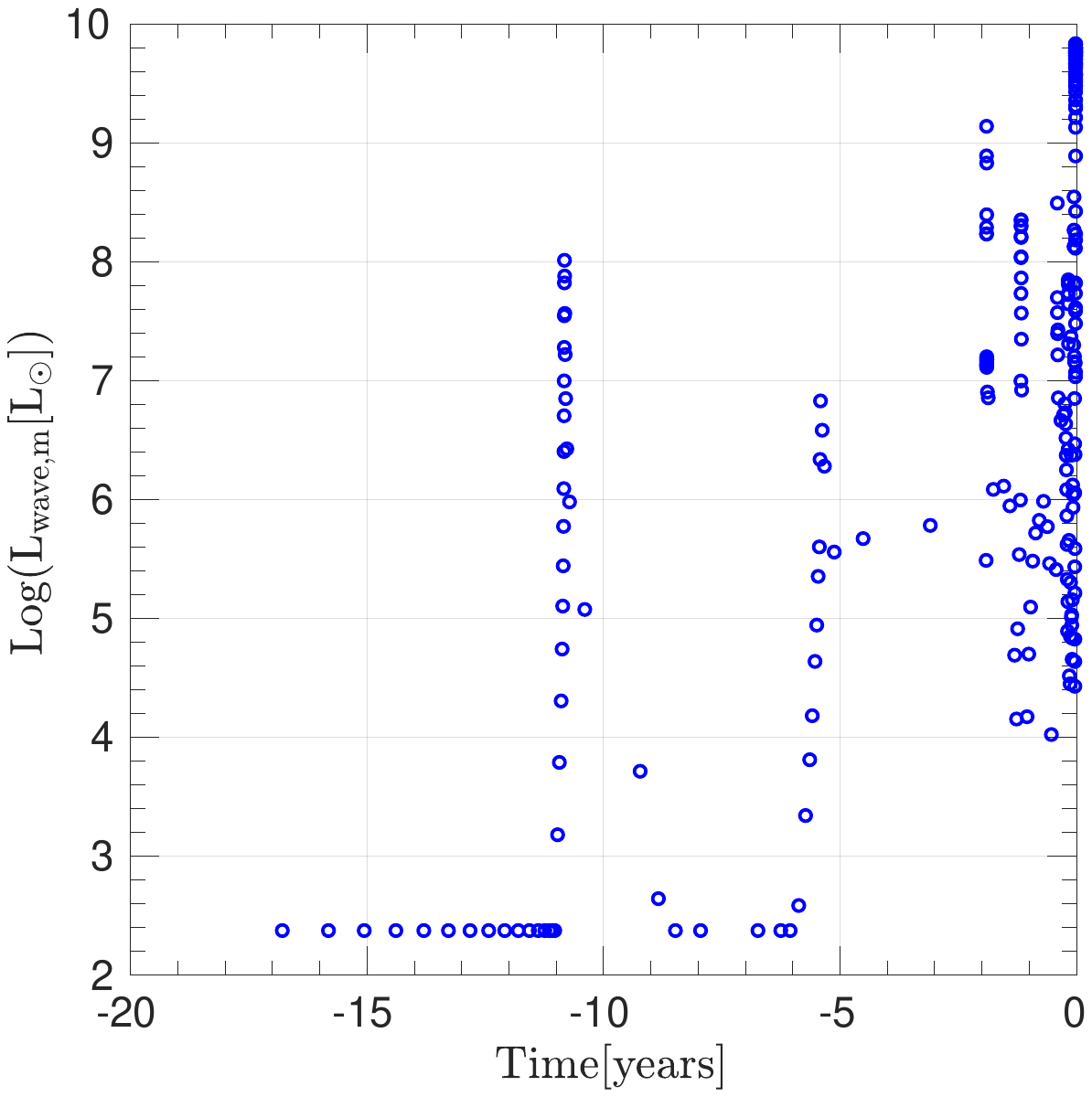}
 \caption{The maximum wave power in the core due to convection in the years before core collapse (at $t=0$). We stop the simulation at specific times to record the maximum wave power (by equation \ref{eq:LwaveFrac}), hence the discrete points of the graph.   
}
\label{fig:LwaveVsYrBC}
\end{figure}

As with wave activity \citep{QuataertShiode2012, ShiodeQuataert2014}, magnetic activity can also occur during the vigorous burning of carbon and neon. We did not simulate these phases here as in this first study we demonstrate the possible implications of magnetic field amplification (dynamo) in the core and the conversion of this energy to thermal energy in the core. In Figure \ref{fig:LwaveVsYrBC} we present the maximum wave power $L_{\rm wave,m}$ during the 17 years before core collapse (at $t=0$). Namely, we check the maximum value of the wave power in the core according to equation (\ref{eq:LwaveFrac}). 
There is a powerful peak lasting for less than a year at $t \simeq -11 \yr$ due to neon burning. 
The power is large, but the duration is short so even if magnetic activity takes place it is not clear the influence on the envelope will be large. This requires a separate study. 

Another effect that Figure \ref{fig:LwaveVsYrBC} presents is that in the five years prior to core collapse the power of wave activity due to oxygen burning has peaks larger than the power we simulated in this study of $\simeq 10^6 L_\odot$. Namely, the effect of magnetic activity might be larger. We notice peaks of $L_{\rm wave,m}>10^7 L_\odot$ at $t \simeq -2 \yr$, $t \simeq -1.2 \yr$, $t \simeq -5 ~{\rm months}$, $t \simeq -2~{\rm months}$, and two short and powerful peaks within less than a month before core collapse (the last one due to silicon burning). On the other hand, if we overestimate the magnetic activity power relative to core convection power, then an influential activity will take place only during the powerful peaks. These might lead to one to three short outbursts in the years before the core collapse (before the explosion). These processes also deserve a separate study. 

Our findings,  that magnetic activity under our assumptions might lead to envelope expansion and luminosity increase, have the following implications. 
\begin{enumerate}
\item The activity we simulated leads to envelope expansion and luminosity increase. These enhance mass loss rates but do not cause an outburst. An outburst most likely requires that these trigger binary interaction (e.g., \citealt{McleySoker2014, DanieliSoker2019}; see section \ref{sec:intro}).  
\item The magnetic activity has cycles, namely, not at all times the dynamo operates at high efficiency. As well, the efficiency of the dynamo depends on the rotation rate of the core, being higher for more rapidly rotating pre-collapse cores (e.g., \citealt{SokerGilkis2017}). This implies that not all CCSNe experience PEOs due to magnetic activity.
\item Even if there is no influence on the envelope due to weak magnetic activity, the dynamo operates at some level (e.g., \citealt{Leidietal2023} for a very recent study). Therefore, in all cases, we expect magnetic activity in the core. \cite{Peresetal2019} argue that the steep entropy rise prevents the buoyancy of magnetic flux tubes; a result that we have used here. As not all magnetic energy is channeled to thermal energy as we simulated in this study, \cite{Peresetal2019} argue that the radiative zone above the iron core stores magnetic energy that collapses with the core. Magnetic fields in the collapsing core play a crucial role in the jittering jets explosion mechanism (e.g., \citealt{Soker2018magnet, Peresetal2019}). Our study adds to the possible connection between PEOs and the explosion mechanism of CCSNe in the frame of the jittering jets explosion mechanism.   
\end{enumerate}


\section*{Acknowledgements}

 We thank an anonymous referee for helpful comments. 
This research was supported by a grant from the Israel Science Foundation (769/20).

\section*{Data availability}
The data underlying this article will be shared on reasonable request to the corresponding author.  


\label{lastpage}
\end{document}